# Sunspot Catalogue of the Valencia Observatory (1920–1928)

V.M.S. Carrasco[a,b], J.M. Vaquero[b], A.J.P. Aparicio[b], and M.C. Gallego[b]

[a] Centro de Geofísica de Évora, Universidade de Évora, Portugal
[b] Departamento de Física, Universidad de Extremadura, Spain

**Abstract:** A sunspot catalogue was maintained by the Astronomical Observatory of Valencia University (Spain) from 1920 to 1928. Here we present a machine-readable version of this catalogue (OV catalog or OVc), including a quality control analysis. Sunspot number (total and hemispheric) and sunspot area series are constructed using this catalogue. The OV catalog's data are compared with other available solar data, demonstrating that the present contribution provides the scientific community with a reliable catalogue of sunspot data.

1. Introduction

Retrieval of information corresponding to past solar activity is fundamental for both a better understanding of the Sun's behaviour and evolution, and to study the effects of the solar cycle on the Earth's climate system (Hoyt and Schatten, 1998; Vaquero and Vázquez, 2009; Usoskin, 2013). When, as a researcher, one can retrieve sunspot data, it is usually of the sunspot number (Vaquero et al., 2011, 2012; Carrasco et al., 2013; Leussu et al., 2013). But occasionally one can even recover sunspot positions (Arlt, 2009; Arlt et al., 2013; Casas and Vaquero, 2014).

The Astronomical Observatory of Valencia University, also known as the Observatorio de Valencia (henceforth, OV), is located in the city of Valencia in the east of Spain. It was founded in 1909 by the Professor of the Faculty of Sciences Ignacio Tarazona y Blanch, who was appointed as its first director. In 1924, Dr Tarazona was succeeded by Juan Antonio Izquierdo Gómez, and later, in 1926, by Vicente Martí Ortells. The Observatory developed a program of monitoring solar activity. The astronomer in charge of the sunspot observations was Tomás Almer Arnau.

The OV's equipment was excellent. It included a 152 mm aperture Grubb refractor telescope, a 110 mm aperture Zeiss telescope, a meridian circle constructed by Mouronval, a theodolite for measuring angles, a photographic camera, a spectroscope, a chronometer for sidereal time, and an astronomical pendulum for mean time, inter alia. This first stage of the Observatory terminated in 1932 with its destruction in a fire at the University. The Grubb refractor was one of the few instruments recovered. The Observatory did not return to normal activity until 1946 (Rivas Sendra, 2009).

The objective of the present work is to analyse the Valencia Observatory's published sunspot catalogue (henceforth, OVc), and provide the scientific community with a machine-readable version. Note that most solar data acquired before the second half of the 20th century have yet to be digitized, which makes it impossible in practice to exploit them scientifically (Lefevre and Clette, 2014; Casas and Vaquero, 2014).

Sunspot catalogues are of particular interest not only because of the information they contain on the number of groups and sunspots, but also because they usually include other important parameters (e.g., as in the present case, the area and the heliographic coordinates). Although several sunspot catalogues are available – with some, such as the Royal Greenwich Observatory (RGO) catalogue, even providing very long time series – they nevertheless present certain limitations, e.g., in their temporal coverage or their lack of mutual overlap (Lefevre and Clette, 2014).

Section 2 describes the OV catalogue. In Section 3, we calculate and analyse the sunspot number from the OVc data, and study the sunspot types in Section 4. Section 5 describes the construction of the butterfly diagram. Section 6 compares the OV and RGO catalogues, and Section 7 compares the sunspot areas recorded in Valencia with those reported by Balmaceda et al. (2009). Finally, the conclusions are drawn in Section 8.

## 2. Sunspot Catalogue

The OV compiled and published a sunspot catalogue during the period 1920–1928, with a gap in the years 1921 and 1922. Figure 1 shows one page of the original published catalogue (Observatorio de Valencia, 1930, page 15). The annual numbers of observation days for the seven years in this time interval were, chronologically, 202, 286, 292, 268, 300, 269, and 276. This corresponds to a temporal coverage of 74.0% of all the days in the seven years of study. The original OVc was published as eight booklets (Observatorio de Valencia, 1928a, 1928b, 1929a, 1929b, 1930, 1931a, and 1931b). The OV sunspot observations were performed using photography. Images of the Sun with a diameter of 10 cm were taken every clear-sky day. The complete process was described by Observatorio de Valencia (1928a,b).

| 1926 | N | Φ B(+) A(−) | $L_c$ W(+) E(−) | $L_0$ | $0^2$ | $10^{-6}$ | T |
|---|---|---|---|---|---|---|---|
| Febrero 5 11h 11m b | 5 | 19,0<br>23,0 | 46,0<br>38,5 | 290,5<br>283,0 | 1,2<br>1,7 | 55<br>76 | I |
| | 17 | 12,0 | 30,0 | 274,5 | 2,0 | 95 | I |
| 6<br>12   23<br>b | 5 | 19,0<br>23,0 | 60,0<br>52,0 | 290,7<br>282,7 | 0,5<br>1,0 | 23<br>45 | I |
| | 17 | 13,0 | 44,0 | 274,7 | 2,0 | 95 | I |
| | 18 | 24,5<br>26,0<br>26,0 | 46,5<br>50,0<br>52,0 | 184,2<br>180,7<br>178,7 | 0,2<br>0,3<br>0,3 | 9<br>13<br>13 | I |
| | 19 | 39,5<br>38,8 | 22,3<br>24,2 | 208,4<br>206,5 | 0,2<br>0,3 | 7<br>11 | I |
| 7<br>9   46<br>r | 5 | 19,5<br>19,5<br>21,2<br>23,0 | 80,0<br>72,0<br>70,0<br>64,0 | 299,0<br>291,0<br>289,0<br>283,0 | 0,3<br>0,5<br>0,3<br>0,5 | 14<br>23<br>14<br>22 | I |
| | 17 | 12,5 | 56,0 | 275,0 | 1,0 | 47 | I |
| | 18 | 25,5<br>22,0 | 36,0<br>35,0 | 183,0<br>184,0 | 0,2<br>0,2 | 9<br>9 | I |
| | 19 | 40,0 | 10,0 | 209,0 | 0,3 | 11 | I |
| | 20 | 20,0 | 82,0 | 137,0 | 4,0 | 182 | IVa |
| | 21 | 19,0<br>19,0 | 78,0<br>84,0 | 141,0<br>135,0 | 0,5<br>4,0 | 23<br>183 | IIb |
| 8<br>13   15<br>m | 20 | 20,0 | 66,0 | 137,9 | 4,0 | 182 | IVa |
| | 21 | 19,0<br>19,0 | 69,0<br>83,0 | 134,9<br>120,9 | 4,0<br>2,0 | 183<br>92 | IIa |
| 10<br>10   38<br>b | 20 | 20,0 | 40,5 | 138,5 | 4,0 | 182 | IVa |
| | 21 | 19,0<br>18,5<br>17,0<br>18,0<br>19,0<br>18,0<br>18,0 | 43,0<br>50,0<br>50,0<br>52,5<br>54,0<br>55,0<br>59,0 | 136,0<br>129,0<br>129,0<br>126,5<br>125,0<br>124,0<br>120,0 | 8,0<br>1,0<br>0,3<br>0,5<br>0,3<br>4,0<br>14,0 | 366<br>46<br>14<br>23<br>14<br>185<br>647 | V |
| 11<br>10   39<br>b | 22 | 22,5<br>22,0<br>22,0 | 23,0<br>21,0<br>19,5 | 188,8<br>186,8<br>185,3 | 1,7<br>0,2<br>0,2 | 76<br>9<br>9 | I |

*Figure 1. An example page of the published catalogue (Observatorio de Valencia, 1930, page 15).*

We have prepared a machine-readable version of the original OVc (see the Supplementary Material to the electronic version of this article). Some lines extracted from our machine readable version are listed as example in Table 1. The format of our version is:

a) The first five columns give the date (the first column the year, the second the month, and the third the day) and time (the fourth column gives the hour and the fifth the minute) of the observation.

b) The sixth column indicates the quality of the observation (*b*, *m*, and *r* for good, poor, and fair observations, respectively).

c) The seventh column gives the group number defined by OV to which each sunspot belongs. A specific group has the same number for different days.

d) The eighth and ninth columns give the heliographic coordinates of the spots (heliographic latitude and longitude, respectively).

e) The tenth and eleventh columns give the area of the sunspots in square degrees of the projection on the solar disk and in millionths of the visible hemisphere of the Sun, respectively.

f) Finally, the twelfth column indicates the classification of the sunspots groups from OV into types following the nomenclature established by Cortie (1901). The type of a same group can vary depending on its evolution.

Table 1: Some lines extracted from our machine readable version.

| Year | Month | Day | Hour | Minute | Quality | N | Lat | Long | S_deg | S_mill | Type |
|---|---|---|---|---|---|---|---|---|---|---|---|
| 1923 | 6 | 20 | 11 | 24 | b | 20 | -3.5 | -13.5 | 0.6 | 29 | I |
| 1926 | 5 | 26 | 10 | 10 | r | 69 | -17.5 | 73 | 5 | 231 | IV a |
| 1928 | 5 | 11 | 11 | 47 | m | 67 | -18 | 34.5 | 3 | 139 | III b |

It is important to note at this point that we performed a straightforward examination of these parameters to detect possible errors in the original catalogue. The objective was not to improve the accuracy of the observations, but to identify and correct any existing mistakes. First, the data for year, month, day, hour, minute, and heliographic longitude and latitude should lie within the logical intervals 1920–1928, 1–12, 1–31, 0–24, 0–60, -90°–+90°, and -90°–+90°, respectively. Second, heliographic longitude for any given group from OV should in general increase by between 5° and 20° per day. And third, since heliographic latitude should remain without major changes over time, we set a range of ±5° for latitude changes of any group registered in OV. Data that did not meet these constraints were examined further and corrected if possible. This led to various errors being identified, most of which were correctable. There remained, however, two cases concerning heliographic longitude due to bad printing of the catalogue in which we could not make the above corrections. Furthermore, other errors have been identified: twenty were due to omission of parameters (six cases with omission of areas, six of groups, and eight of type) and the two remained cases were due to bad printing of the respective catalogue in the section 'Type'. We did not count as errors the many cases in which there was no information as to whether the observation was good, poor, or fair. These twenty-four errors represented 0.19% of the total OVc records. Omitted groups were located visually.

## 3. Solar Indices

### 3.1. Sunspot number

Sunspot numbers were calculated following the definitions of the International Sunspot Number (ISN) and Group Sunspot Number (GSN) reference indices (Hoyt and Schatten, 1998; Vaquero, 2007; Clette et al., 2007). We thus define the Valencia International Sunspot Number (VISN) as VISN = 10 $g+f$ and the Valencia Group Sunspot Number (VGSN) as VGSN = 12.08 $g$, with $g$ being the number of groups and $f$ the number of individual spots observed in the OV. Figure 2 shows the evolution of these two series of OVc indices, and their comparison with the ISN (http://sidc.oma.be/silso/datafiles) and GSN (http://www.ngdc.noaa.gov/stp/space-weather/solar-data/solar-indices/sunspot-numbers/group/) indices.

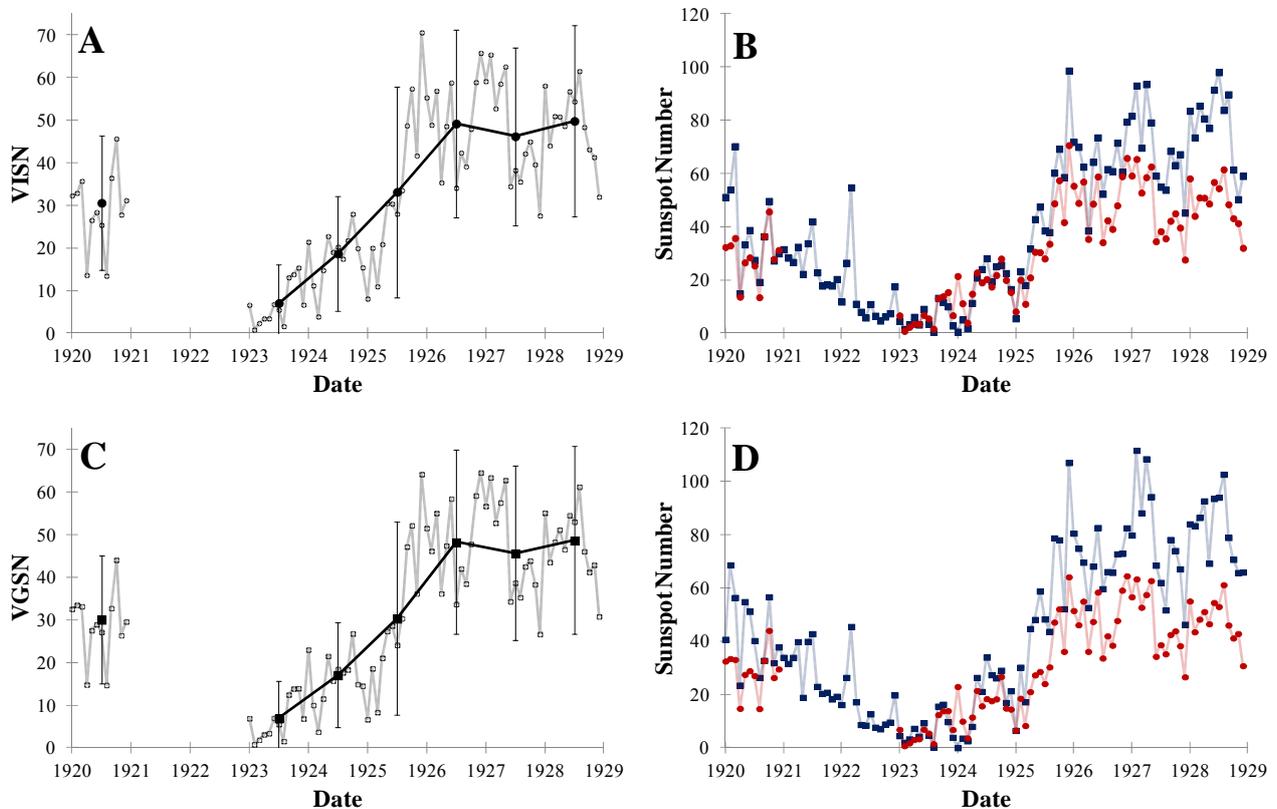

*Figure 2. (A) Monthly (open discs) and annual (solid discs) values of VISN. (B) Monthly ISN (blue*

*squares) and VISN (red discs) data. (C) Monthly (open squares) and annual (solid squares) values of VGSN. (D) Monthly GSN (blue squares) and VGSN (red discs) data.*

The OV registered a significant rise in solar activity in the second half of 1925 (Figures 2A,C). There was a fall in 1927, followed by a second rise in early 1928, producing a Gnevyshev gap (Gnevyshev, 1967, 1977). The behaviour of the VISN and VGSN indices is similar except that the annual VISN values are slightly higher. One observes in Figures 2B,D that the reference indices are generally higher than the corresponding Valencia Sunspot Numbers. The exceptions are those in late 1923 and early 1924, when the Valencia indices are significantly higher than both of the reference values. In general, the difference between the Valencia and the reference data is greater for the GSN than for the ISN case.

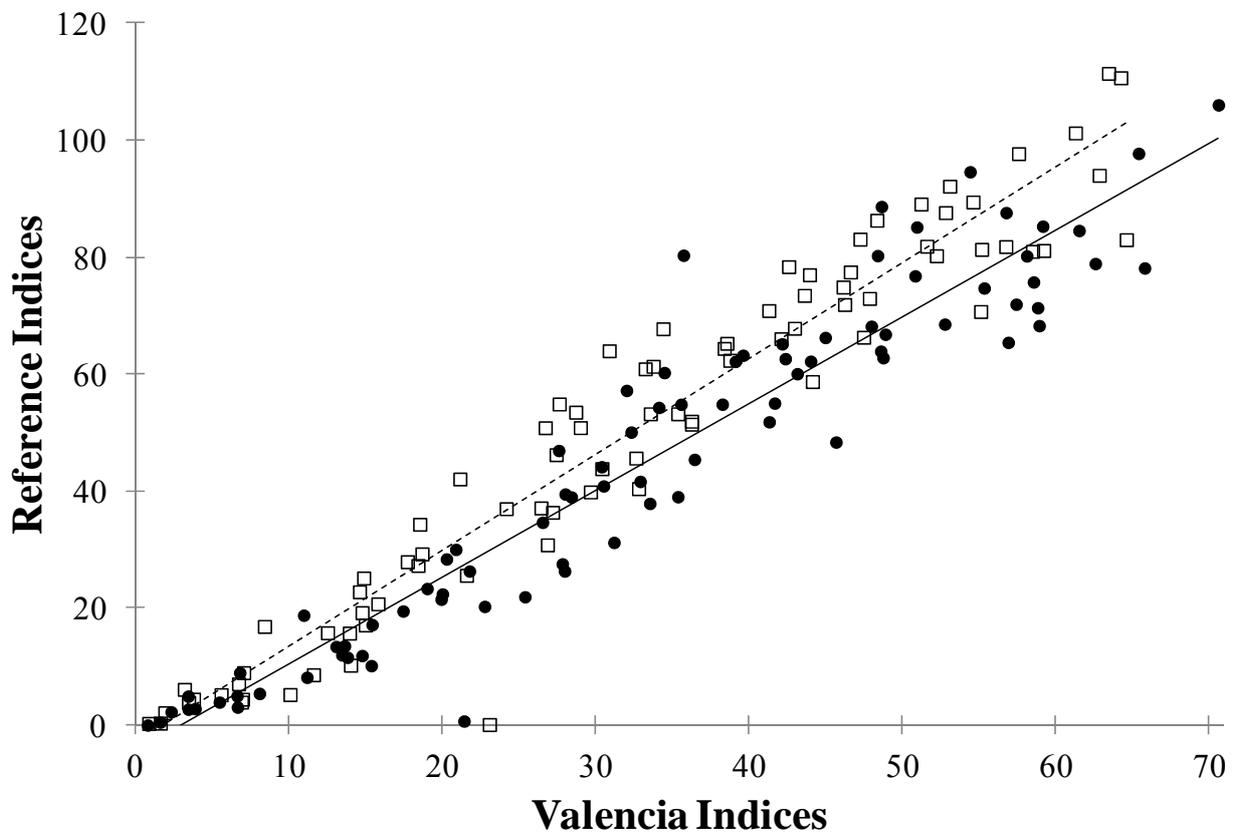

*Figure 3. Linear correlations: ISN-VISN (discs) and GSN-VGSN (squares).*

We obtained a calibration factor for the OV data by correlating them with the corresponding data of the two reference series (Figure 3) for days with data for both indices. In particular, we found the best linear fits to be: ISN=(1.48±0.05)VISN+(–4.5±2.0); r=0.954; p-value<0.001; and GSN=(1.64±0.05)VGSN+(–3.0±1.8); r=0.965; p-value<0.001. Setting the *y*-intercept of the regression lines to zero, we obtained the calibration factors $k_{VISN}$ = 1.38 ± 0.03 and $k_{VGSN}$ = 1.57 ± 0.02.

To study the variability of the indices, we examined the annual VISN/VGSN ratios as calculated from the monthly data (Figure 4), and compared them with the corresponding reference ratios (ISN/GSN). One observes in the figure that the two OV series are robust in the sense that the evolution of their ratio presents no marked variations. Indeed, while the Valencia ratios are greater than those of the reference series, their standard deviations are smaller. We must note that points for which the value of any ratios was equal or less than 3 were excluded from these calculations.

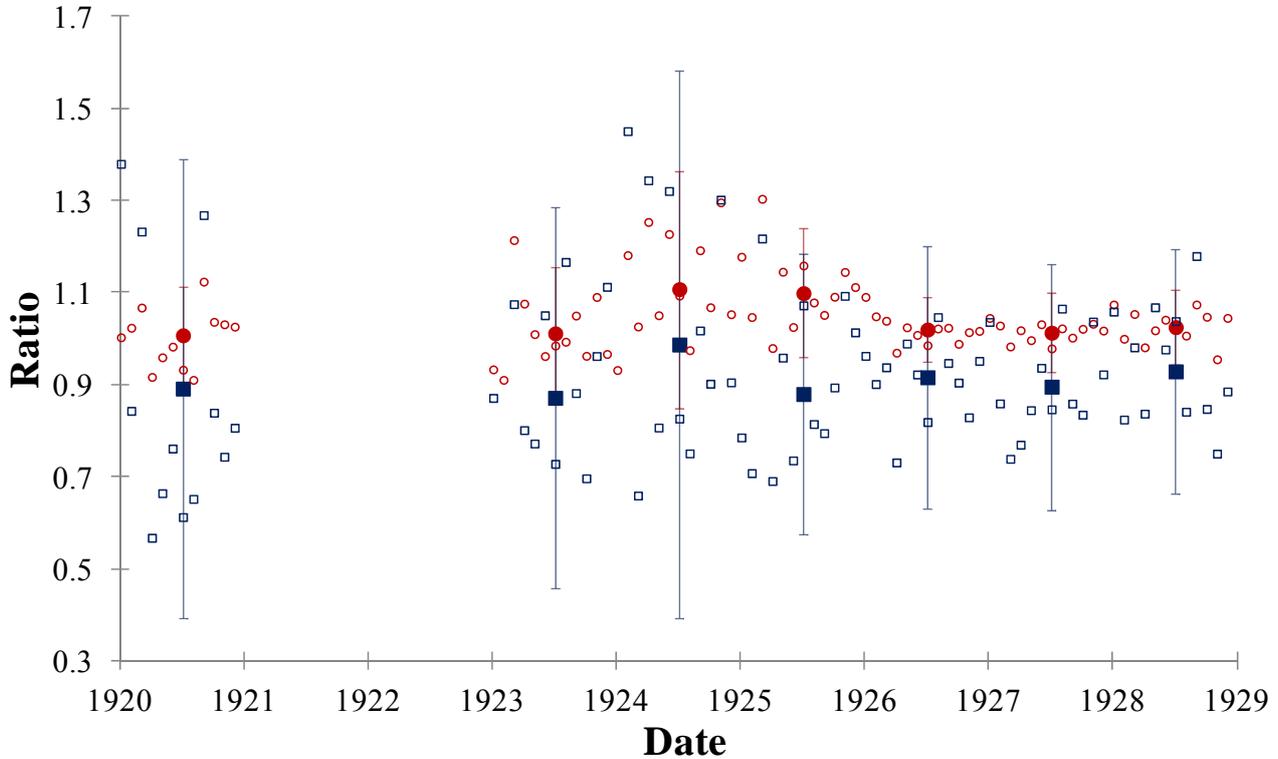

*Figure 4. The ratios VISN/VGSN (discs) and ISN/GSN (squares). The error bars represent plus and minus one standard deviation.*

*3.2. Hemispheric sunspot number*

For the calculation of the two hemisphere indices that we shall term the Northern/Southern Valencia International Sunspot Number ($VISN_N$/$VISN_S$) and the Northern/Southern Valencia Group Sunspot Number ($VGSN_N$/$VGSN_S$), we applied the definitions given in Section 3.1 to the two hemispheres (north and south). Figure 5 shows plots of the monthly and annual values of these indices.

Figure 5 shows how the solar activity of the complete cycle is the composition of the two hemispheres' activities (VGSN and VISN). The maxima of the annual values do not coincide in time in the two hemispheres. In the northern hemisphere, the maximum occurs in 1928. In the south, it occurs a year earlier, in 1927. For both hemispheres, the minimum occurs in 1923. The fall that one observes in the northern hemisphere solar activity in early 1927 is responsible for the formation of the Gnevyshev gap at around that date. Zolotova et al (2009, 2010) showed from RGO data that two significant changes in the predominant hemispheric leading have been detected since 1874. The northern hemisphere was phase-advanced until near 1928 when the first change occurred. From both hemispheres indices (Figure 5), we can see that a change in the phase-advanced hemisphere occurred around the years 1927-1928 (around the maximum of solar cycle No. 16).

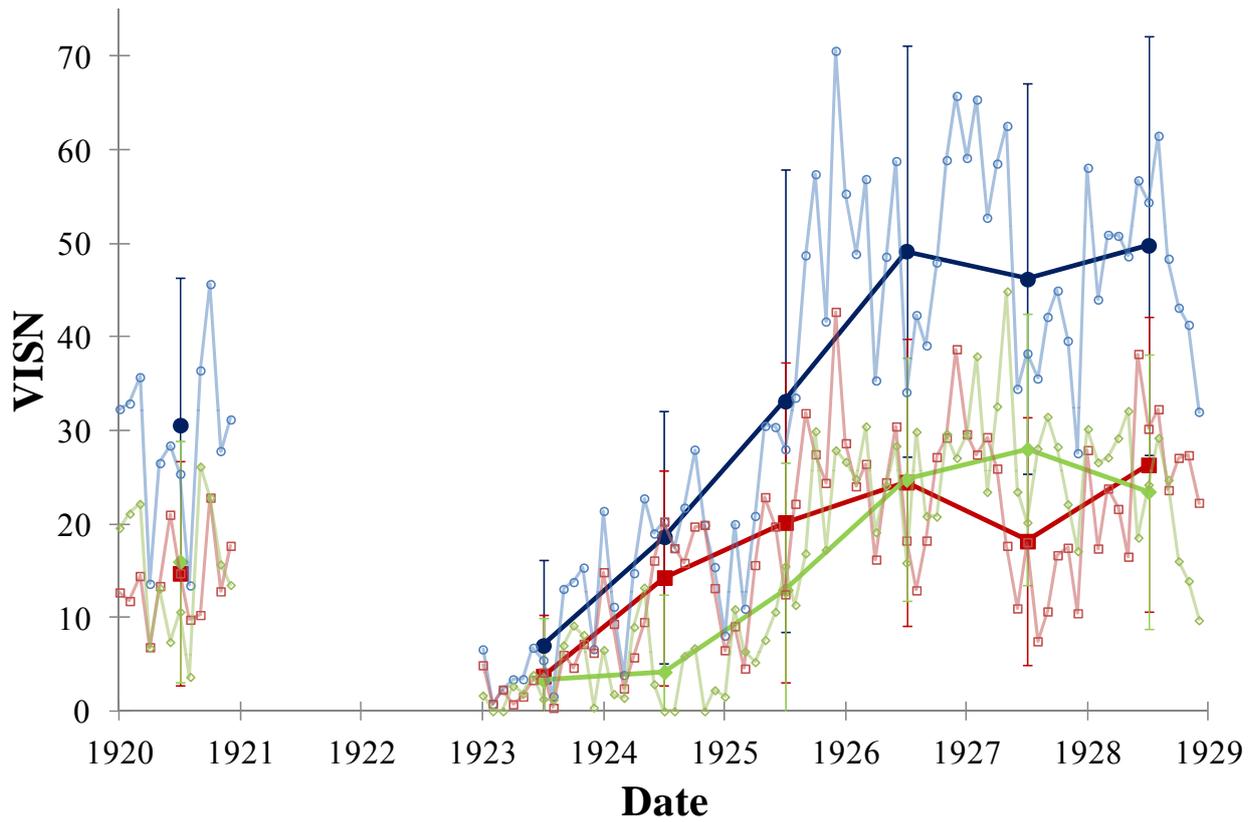

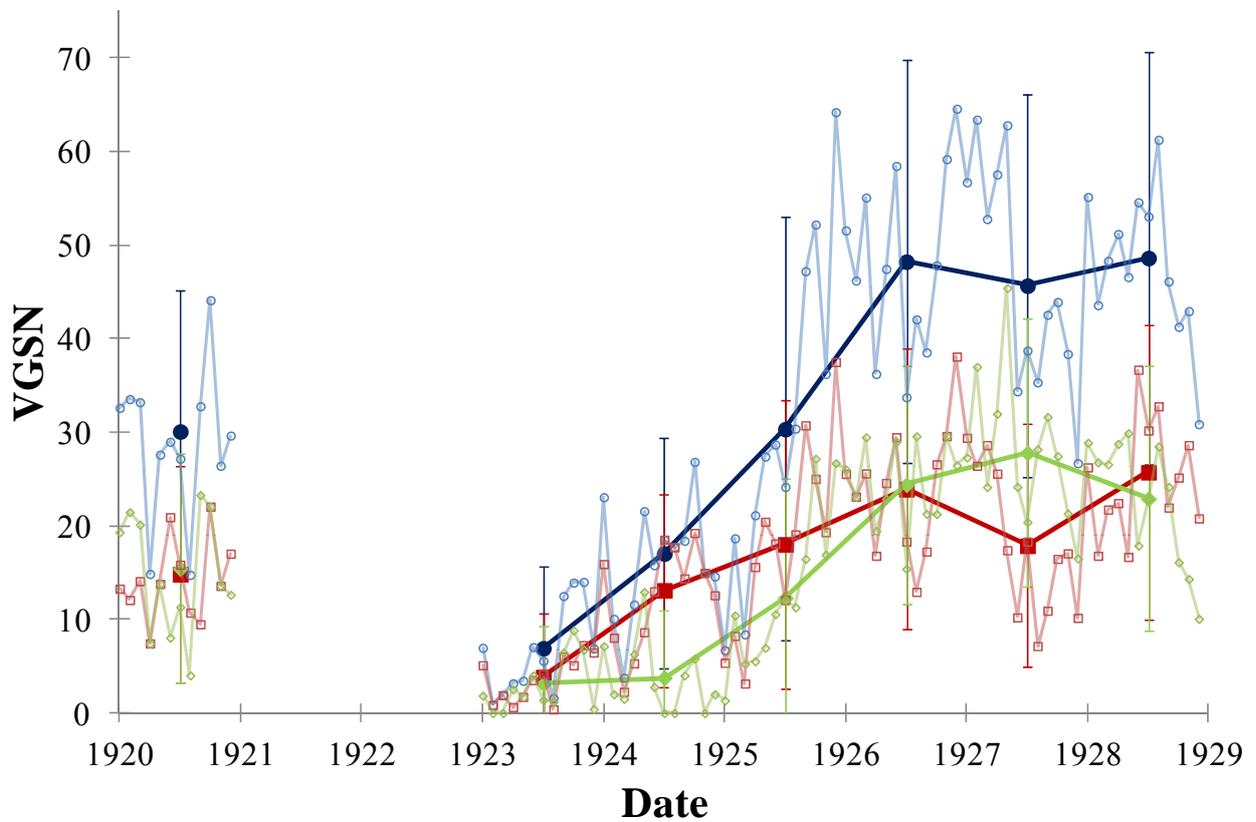

*Figure 5. The upper/lower panels are plots of VISN/VGSN (blue), VISN$_N$/VGSN$_N$ (red), and VISN$_S$/VGSN$_S$ (green). The solid discs represent annual values, and the open discs monthly values. Error bars represents one standard deviation.*

*3.3. Normalized asymmetry*

Various workers have studied the north-south asymmetry of the solar activity cycle using the data of the available solar activity indices (Carbonell, Oliver, and Ballester, 1993; Berdyugina and Usoskin, 2003; Joshi and Joshi, 2004). There is some difference in the behaviour that these studies report according to whether an absolute asymmetry ($A = R_N - R_S$, where R is a solar activity index) or a normalized asymmetry (Temmer et al., 2006) is used. In the present study, we used the normalized asymmetry (NA) of GSN: $NA = (R_{GN} - R_{GS})/(R_{GN} + R_{GS})$, where $R_G$ is the GSN index and the subscripts *N* and *S* refer to the northern and southern hemispheres, respectively. Figure 6 is a plot of the values of this normalized asymmetry as calculated from the OVc data.

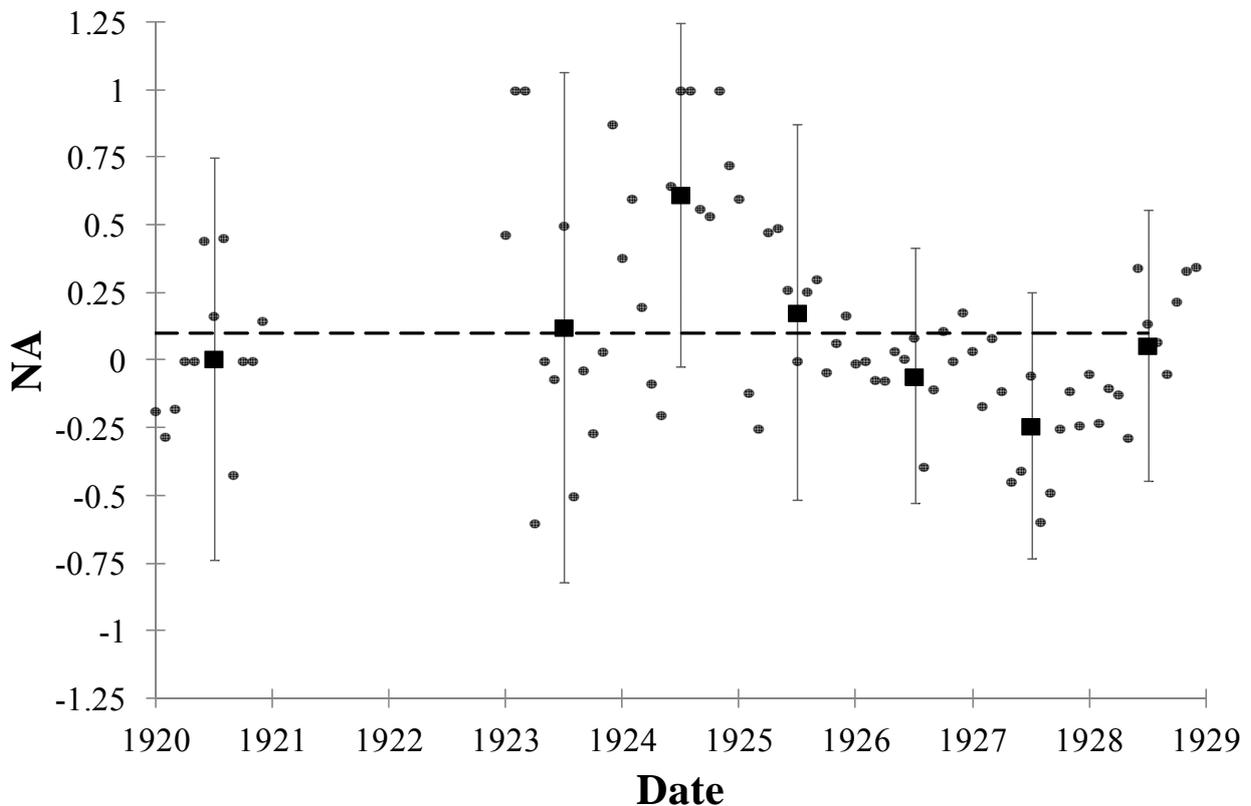

*Figure 6. Temporal evolution of the normalized asymmetry, NA. The squares (circles) represent annual (monthly) values. The dashed line represents the mean annual value of NA for the entire period recovered. The error bars represent plus and minus one standard deviation.*

The mean annual NA (represented by the dashed line in Figure 6) is positive (0.097). The northern hemisphere therefore had on average for Solar Cycle 16 a slightly greater weight in the overall solar activity as registered by the Valencia observations than the southern hemisphere. But one also observes in the figure that, for this solar cycle, while the northern hemisphere was the dominant hemisphere around the minimum of solar activity, around the maximum, it was the southern hemisphere.

## 4. Sunspot Types

OVc classifies the types of sunspots on the basis of the original proposal of Cortie (1901). This classification (Figure 7) distinguishes five principal types of sunspots: Type I comprises groups with one or more sunspots of small size; Type II refers to formations of two sunspots in which one or both may be principal; Type III refers to trains of spots; Type IV refers to individual spots with determined shapes; and Type V comprises irregular groups of large sunspots (Figure 7).

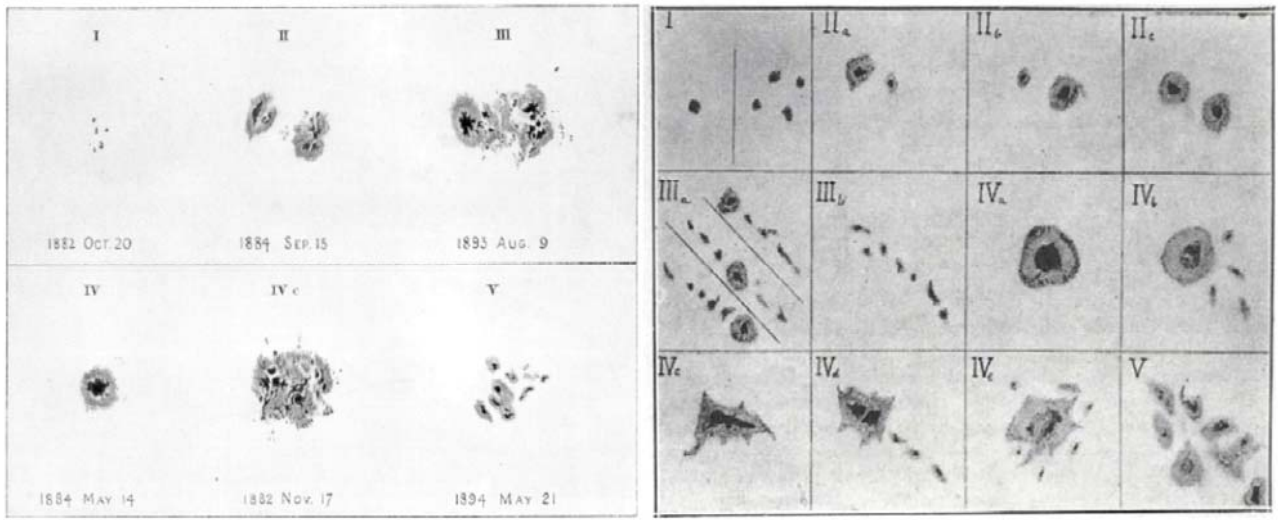

*Figure 7. Cortie's classification of sunspots. [Sources: (left) Cortie (1901); (right) Observatorio de Valencia (1928a)].*

Figure 8 shows the percentages corresponding to the different types of sunspots recorded in the OV for the separate years of the catalogue and for the entire period.

Type I accounts for the largest fraction of sunspots. It is a type which represents the initial and final phases of the formation of groups. Nevertheless, in 1920, the fraction represented by the sum of all the Type IV subtypes exceeds that of Type I, and both of these types account for greater percentages than the others in this catalogue for every year. The spotless days (SD) recorded represent 4.7% of all observations, while the sunspot cases without a type being determined (WT) represent 0.2% of the total.

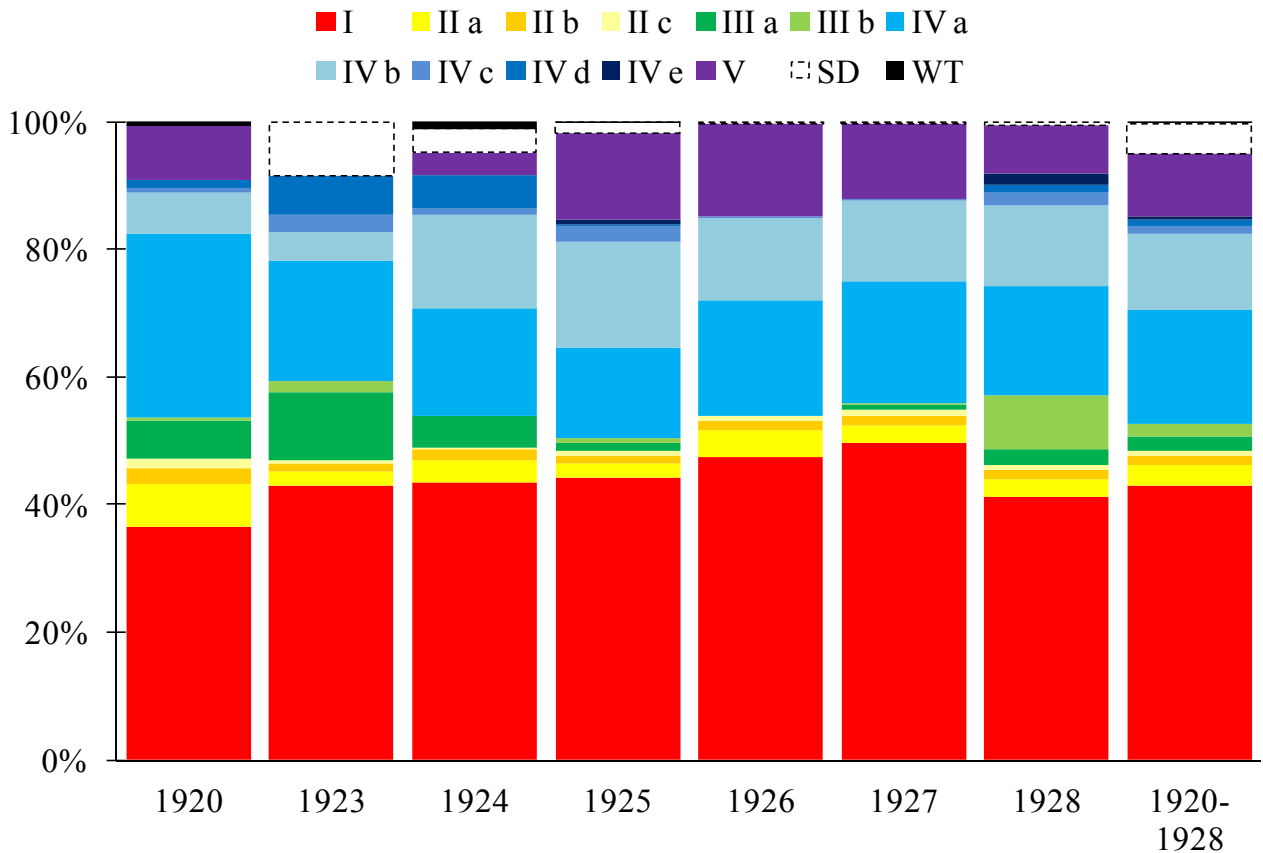

*Figure 8. Study of the percentage of the different types of sunspot for entire period and each of the years recovered from OVc. The legend indicates the colours assigned to each of the types of sunspot groups, where SD represents spotless days and WT days without classifications,* i.e. *with no sunspot types determined.*

## 5. Butterfly Diagram

To gain some insight into the temporal evolution of the distribution of the latitudes at which the sunspots were observed, we constructed a butterfly diagram (Figure 9) from the OV data. The diagram only covers latitudes between ±40° because no spots in the catalogue lay outside this range (according to RGO data).

One observes in this butterfly diagram that solar activity at the start of Solar Cycle 16 is centred primarily at latitudes near ±20°. As was to be expected, the sunspots form at lower latitudes as the cycle progresses until, at the end of the cycle, they concentrate towards the vicinity of the solar equator as they also did for the end of Solar Cycle 15. One also observes that the sunspots increase not only in number at the maximum of the cycle, but also in size.

At the beginning of Solar Cycle 16, the spots are greater in number and larger in size in the north than in the south. During the maximum of this cycle however, there are more and even larger spots in the south. In the data retrieved corresponding to the tail of Cycle 15, there are slightly more sunspots in the southern hemisphere, and they are larger in size than those in the northern hemisphere.

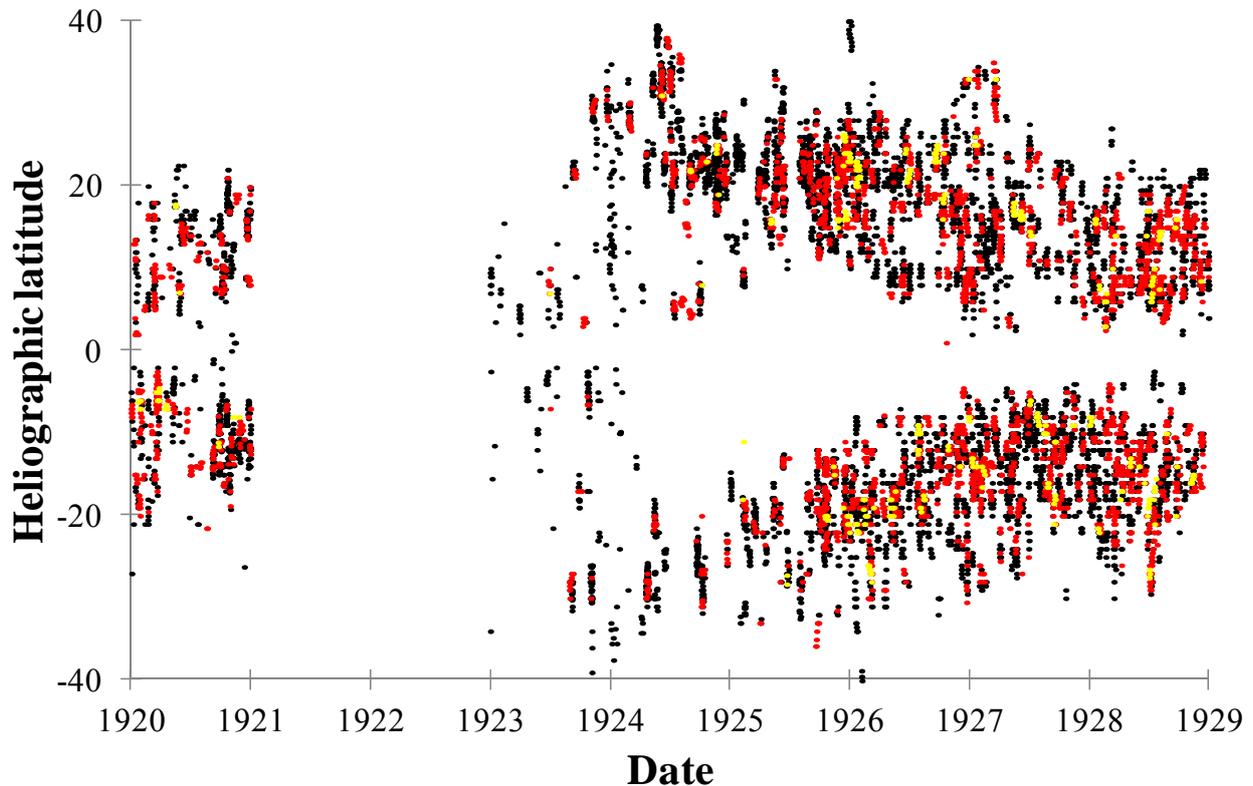

*Figure 9. Butterfly diagram of the OVc sunspot record latitudes. The black, red, and yellow dots represent sunspot areas less than 100 ppm, between 100–499 ppm, and greater than or equal to 500 ppm, respectively.*

## 6. Comparison with the RGO Catalogue

Many studies have used the RGO data as reference. Hathaway and Choudhary (2008), for example, used them to determine the decay rate of sunspot group areas, and Balmaceda et al. (2009) constructed a sunspot area dataset from them. However, isolated, systematic, and typographical errors have been detected in the different published RGO datasets. Recent work has been targeted at identifying and correcting these errors (Willis et al., 2013a; Willis et al., 2013b; Erwin et al., 2013). The aim in this section is to perform a comparison between the data retrieved in this paper from OV and the data registered in RGO. Thus, we obtained the RGO data from the Website http://solarscience.msfc.nasa.gov/greenwch.shtml. Since this database only lists group data, and OVc provides information on sunspots (including the group number to which each spot belongs), we can carry out the comparison between data from both observatories after recalculating all parameters in OVc in accordance with its groups. For this purpose, we average each parameter for sunspots associated with the same group.

First, we compared the groups registered at both observatories. Then, before beginning the group correspondence and in order to correctly associate groups in the two catalogues, we imposed constraints on the differences in latitude and longitude. To determine the optimal values for these differences, we made a statistical search for the greatest number of one-to-one correspondences that are found for RGO, i.e., number of groups in OVc with a unique correspondence in RGO (Table 2). Therefore, in this first step we do not study the different situations that are presented to associate groups that will be explained below.

One observes in Table 2 that the interval which gave the greatest number of identifications was that corresponding to differences of ±3° for latitude and ±15° for longitude. The number of one-to-one correspondences found was then 765, i.e. 84.7% of the total number of groups registered in Valencia. This interval was therefore selected as the criterion to use for our identifications.

*Table 2: Number of groups in OV that were uniquely identified in RGO for different combinations of allowed differences in latitude and longitude.*

| LATITUDE (°) | 1 | | | 2 | | | 3 | | |
|---|---|---|---|---|---|---|---|---|---|
| LONGITUDE (°) | *15* | *20* | *25* | *15* | *20* | *25* | *15* | *20* | *25* |
| **Correspondences** | 755 | 754 | 752 | 764 | 762 | 756 | 765 | 762 | 754 |
| LATITUDE (°) | 4 | | | 5 | | | 7.5 | | |
| LONGITUDE (°) | *15* | *20* | *25* | *15* | *20* | *25* | *15* | *20* | *25* |
| **Correspondences** | 760 | 756 | 744 | 761 | 756 | 744 | 760 | 753 | 740 |

Then, we apply the selected constraints to the OV data. Therefore, a group in RGO is associated with a group at OV if it meets the differences of ±3° in latitude and ±15° n longitude. Cases where we found correspondences on isolated days or different groups are associated with only one group in the other observatory are individually analysed. We should note that 765 is the number of cases identified uniquely without taking into account the several possibilities presented to associate groups, explained below. For example, groups in OV which have several possible corresponding groups in RGO data are not considered in that sum. Table 3 presents the statistics for the different cases in attempting to match groups between the two observatories. The total number of groups recorded for the study period in OV was 906, and in RGO was 2261. Of the OV groups, 762 (84.1%) were identified in RGO. In this case, 762 is the number of the final cases of matches after considering the different options that were presented. Of the RGO groups, 893 (39.5%) were identified in OVc. Some of the cases of matches between the two observatories were where an OV group had a multi-correspondence with RGO groups. On the one hand, individual analysis of the latitude, longitude, and area parameters of these groups showed that some group or groups did not have corresponding groups in the other dataset despite the fact that all met our constraints for acceptance. In particular, there were 133 (14.7%) OV groups with preliminary multi-correspondences that further analysis showed not to correspond to any of the groups that had initially been assigned. On the other hand, 4 (0.4%) OV cases with a preliminary correspondence were found on further analysis to have no counterpart in RGO data. In total, there were of 144 (15.9%) OV groups and 1368 (60.5%) RGO groups with no match found. The number of groups for days on which observations were recorded at both observatories were 848 (93.6%) for OV and 2005 (89.5%) for RGO. Finally, there were 256 RGO groups on the days when no observation was available in OVc, and 58 for the opposite case.

*Table 3: The different cases found in matching the OV and RGO groups.*

| Case | OV | RGO |
|---|---|---|
| Total groups | 906 | 2261 |
| Matching groups | 762 | 893 |
| No match | 144 | 1368 |
| Observations at VO and RGO | 848 | 2005 |
| No VO observation for RGO | - | 256 |
| No RGO observation for VO | 58 | - |

In addition, we evaluated a calibration factor for the OV data with respect to RGO as reference from a comparison between their series, VGSN and GGSN, respectively. Figure 10 (top) shows the linear regression fit between the two observatories' monthly Group Sunspot Number data when both observatories have observations for a same day. The straight line fit is given by the expression: GGSN = $(1.62 \pm 0.06)$ VGSN + $(1.0 \pm 2.2)$, $r = 0.950$, $p$-value < 0.001. Setting the $y$-intercept of the regression line to zero, we obtained the calibration factor for OV with respect to RGO: $k_V = 1.64 \pm 0.03$. The bottom plot of Figure 10 shows the RGO and OV series together with the OV series corrected by the above calibration factor. Error bars represent one standard deviation.

In Figure 10, one notes obvious differences between both series (GGSN and VGSN). One is in the values – GGSN has considerably higher values than VGSN. This difference in the values is due to a larger number of groups in RGO were registered. The principal reason for this fact is due to RGO registered a higher number of small groups than OV. In particular, 1315 groups with an area value lower than 30 millionths of the solar hemisphere were registered in RGO but not in OV (the minimum area reported in OV is 3). It is also noteworthy that sometimes one group in OV is associated with several groups in RGO (68 cases), almost three times more than *vice versa*, approximately. Furthermore, OV assign the same group number for recurrent groups (RGO gives a new number) and it rarely registers groups with heliographic lengths above 80º. Another difference is in the shape of the two curves – while VGSN has a small Gnevyshev gap, GGSN presents continual growth in its ascending phase until reaching the maximum.

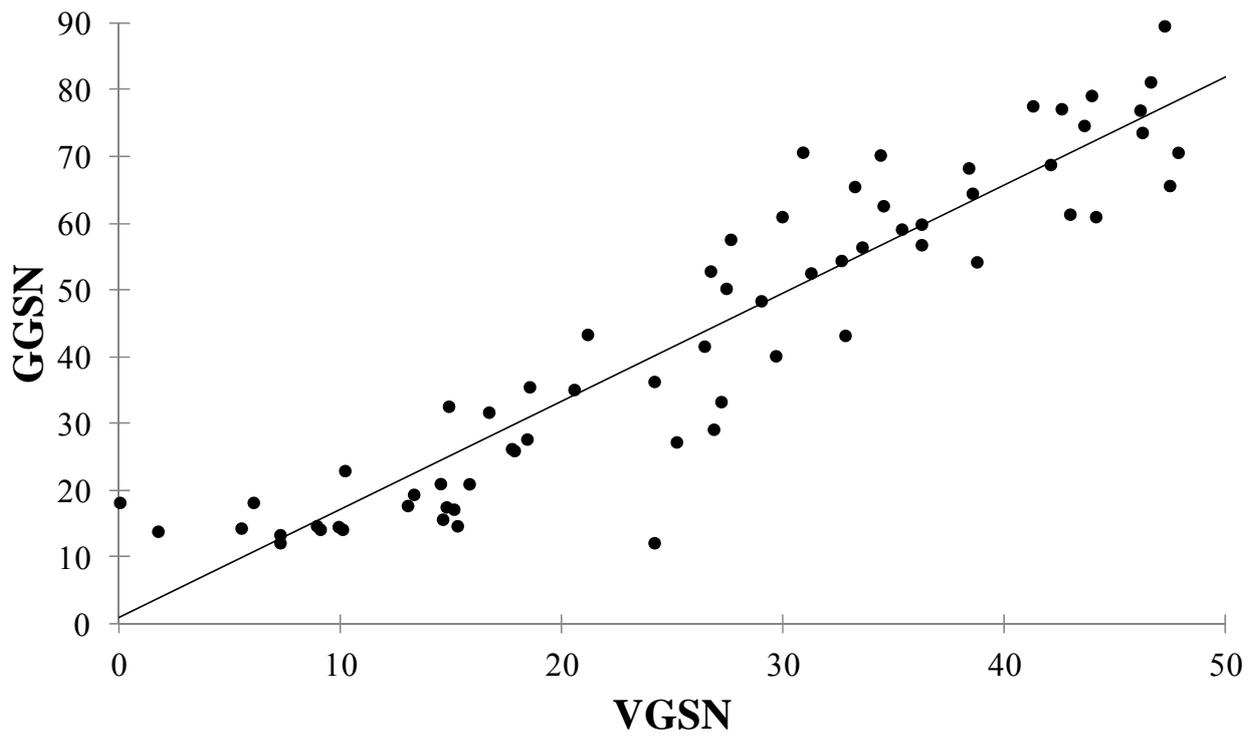
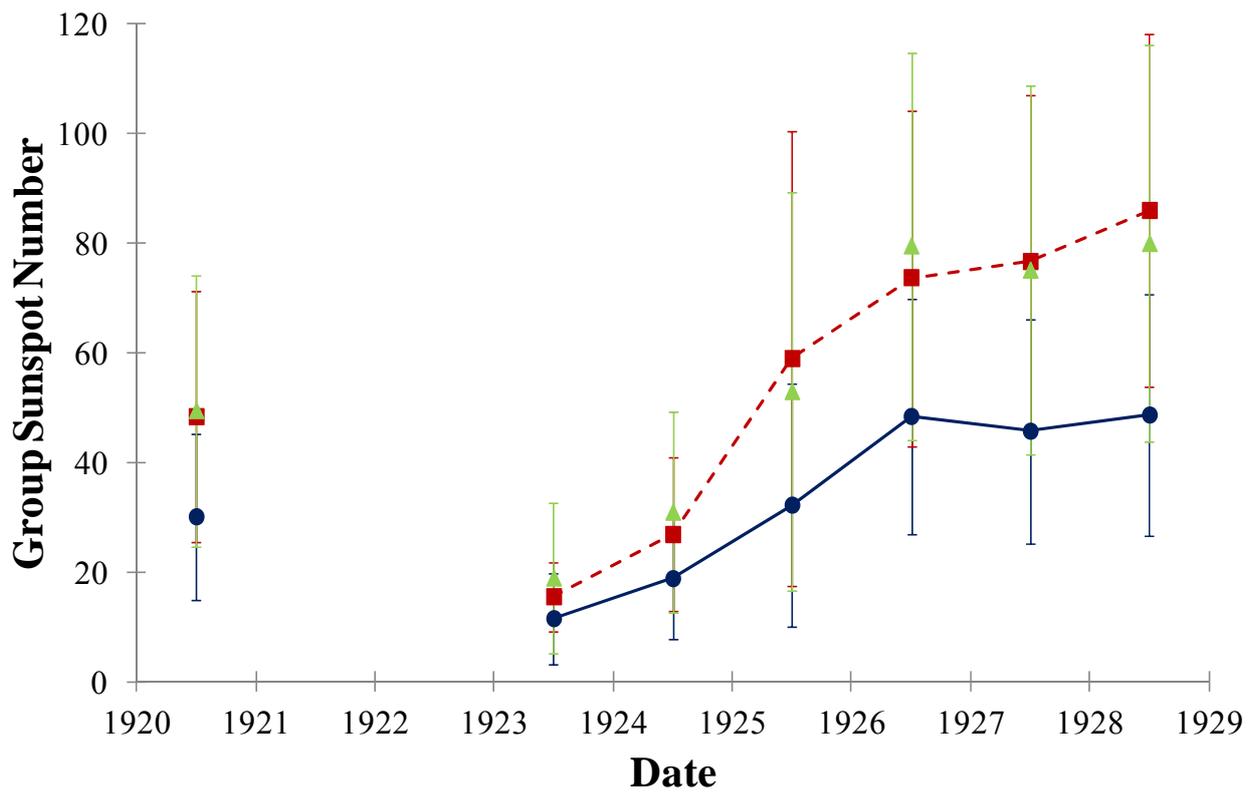

Figure 10. (Top) Linear correlation between GGSN and VGSN. (Bottom) Plot of the GGSN (squares), raw VGSN (discs), and corrected VGSN (triangles) time series; error bars indicate one standard deviation.

## 7. Comparison with the Balmaceda Database

The result of our construction of the Valencia Sunspot Area (VSA) series from the recovered data in the OVc is shown in Figure 11. One observes that, while the difference relative to 1928 is only very slight, the VSA maximum occurs in 1926. The minimum occurs in 1923. While the minimum of the cycle is coherent with those of VISN and VGSN described above, this is not the case for the maximum, since the VISN and VGSN maxima occur in 1928. Moreover, VSA also has a Gnevyshev gap as do VISN and VGSN. With respect to the monthly data, one notes the high solar activity in December 1925 and January 1926, and the relatively low value in December 1927.

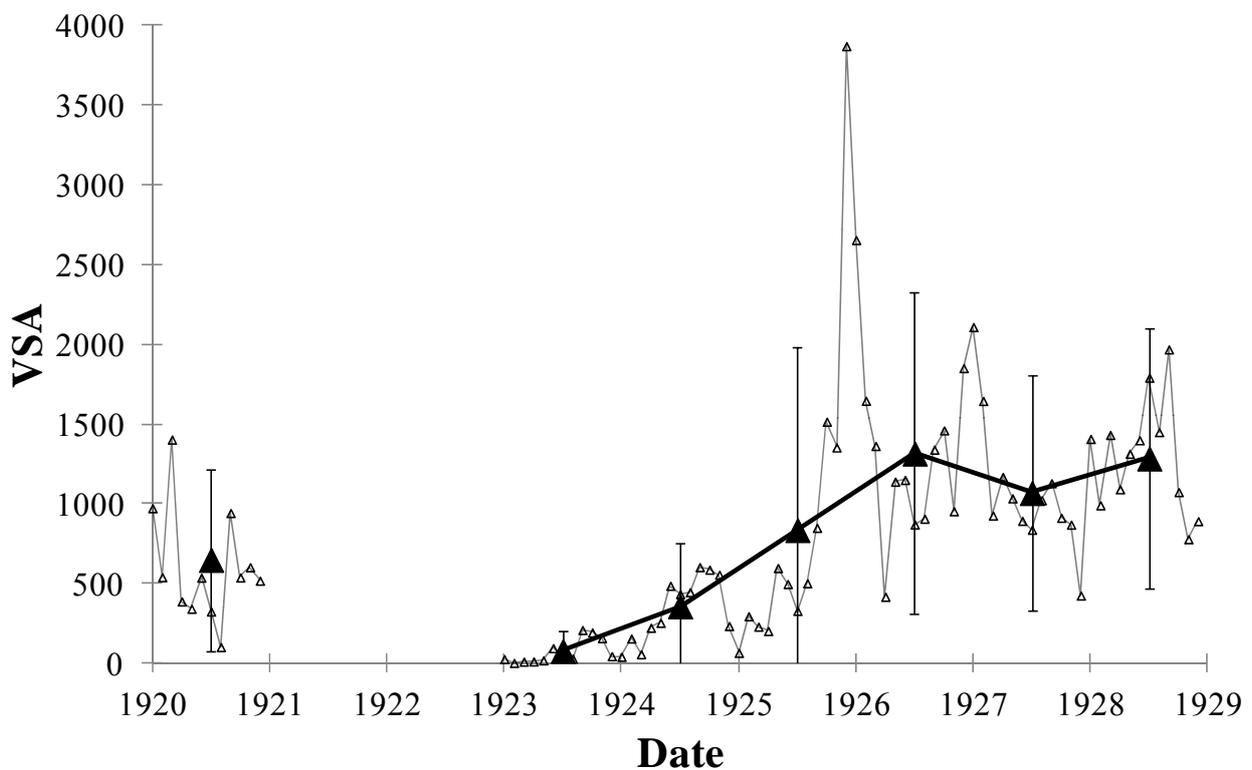

*Figure 11. The monthly (small triangles) and annual (large triangles) Valencia Sunspot Area (VSA) time series. Error bars indicate one standard deviation. The units of the y-axis are millionths of hemisphere.*

We compared the VSA values with the database (BSA) created by Balmaceda et al. (2009) using the latter's methodological approach (Figure 12). For the calculations of Figures 12A,B, we excluded data with a deviation greater than ±3σ and or which were located below the straight line joining the points (0, 3σ) and (3σ, 0). The regression line in Figure 12A is given by the expression BSA = (1.04 ± 0.02) VSA + (-32 ± 22), $r$ = 0.990, $p$-value < 0.001. And that in Figure 12B is given by VSA = (0.94 ± 0.02) BSA + (52 ± 20), $r$ = 0.990, $p$-value < 0.001. Setting the $y$-intercept of the regression lines to zero and averaging the results (the slope is inverted in the second expression), we obtained the calibration factor $k_{VSA} = 1.02 \pm 0.01$.

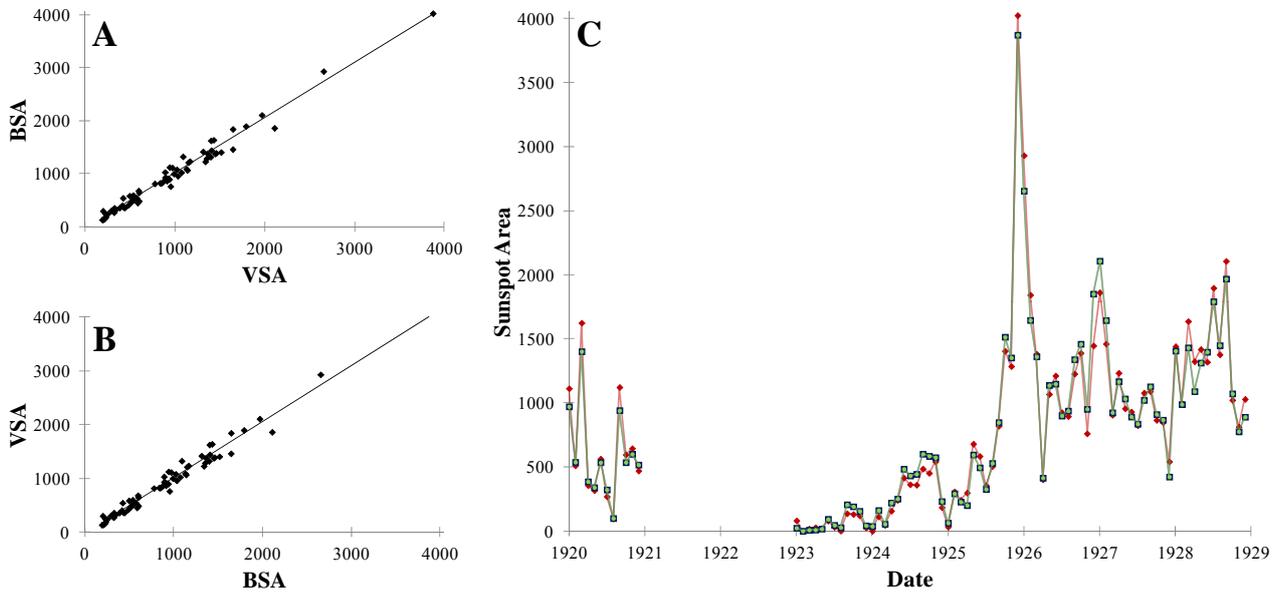

*Figure 12. (A and B) Linear correlation between VSA and BSA. (C) Comparison of the monthly values of BSA (red), VSA (blue), and corrected VSA (green).*

One observes in Figure 12C that all three series present similar behaviour. In general, the VSA values are slightly greater than those of BSA, except for 1928 and the maximum epoch for this solar cycle when the reverse is the case. There stand out the sharp peaks in activity in late 1925 and early 1926 in both datasets, followed by a sharp drop in solar activity. There were eight days of observation with records in OVc but not in the Balmaceda database. The OV observations may be considered as candidates to fill those gaps.

## 8. Conclusions

In this study, we have retrieved and digitized the data corresponding to the printed version of the Valencia catalogue (OVc). In this manner, a new sunspot catalogue is now available in digital format. We carried out a quality control analysis to identify and correct errors. The Valencia Sunspot Numbers were corrected against the ISN and GSN reference indices, and the hemispheric Sunspot Numbers and normalized asymmetry were presented. A morphological classification was given of the OVc sunspots, and the OVc butterfly diagram was presented. Finally, we compared OVc with the RGO and Balmaceda databases. Most of the groups recorded in OV were also identified in RGO, while there being clear differences between the two series. However the Balmaceda and Valencia Sunspot Area series showed similar behaviour, resulting in a calibration factor of 1.02.

**Acknowledgements**

Support from the Junta de Extremadura (Research Group Grant No. GR10131), from the Ministerio de Economía y Competitividad of the Spanish Government (AYA2011-25945) and from the COST Action ES1005 TOSCA (http://www.tosca-cost.eu) is gratefully acknowledged.